\documentclass[10pt,conference]{IEEEtran}
\IEEEoverridecommandlockouts

\usepackage{cite}
\usepackage{amsmath,amssymb,amsfonts}
\usepackage{algorithmic}
\usepackage{graphicx}
\usepackage{textcomp}
\usepackage{tcolorbox}
\usepackage{todonotes}
\usepackage{url}
\usepackage{listings}
\usepackage{booktabs}
\usepackage{xspace}
\usepackage{hyperref}

\lstdefinestyle{mystyle}{
    basicstyle=\ttfamily\footnotesize,
    captionpos=b,                    
    numbers=left,                    
    numbersep=5pt,                  
    tabsize=2
}
\newcommand{\urlRepo}[0]{\url{https://osf.io/nw8vp/?view_only=a4654139dce54584ab5e1549e56dff3c}\xspace} 

\usepackage{xcolor}
\def\BibTeX{{\rm B\kern-.05em{\sc i\kern-.025em b}\kern-.08em
    T\kern-.1667em\lower.7ex\hbox{E}\kern-.125emX}}

\begin{document}

\title{Studying How Configurations Impact Code Generation in LLMs: the Case of ChatGPT
}
\author{\IEEEauthorblockN{Benedetta Donato}
\IEEEauthorblockA{
\textit{University of Milano-Bicocca}\\
Milano, Italy \\
benedetta.donato@unimib.it}
\and
\IEEEauthorblockN{Leonardo Mariani}
\IEEEauthorblockA{
\textit{University of Milano-Bicocca}\\
Milano, Italy \\
leonardo.mariani@unimib.it}
 \and
\IEEEauthorblockN{Daniela Micucci}
\IEEEauthorblockA{
\textit{University of Milano-Bicocca}\\
Milano, Italy \\
daniela.micucci@unimib.it}
\and
\IEEEauthorblockN{Oliviero Riganelli}
\IEEEauthorblockA{
\textit{University of Milano-Bicocca}\\
Milano, Italy \\
oliviero.riganelli@unimib.it}

}

\maketitle

\begin{abstract}
Leveraging LLMs for code generation is becoming increasingly common, as tools like ChatGPT can suggest method implementations with minimal input, such as a method signature and brief description. Empirical studies further highlight the effectiveness of LLMs in handling such tasks, demonstrating notable performance in code generation scenarios.

However, LLMs are inherently non-deterministic, with their output influenced by parameters such as temperature, which regulates the model's level of creativity, and top-p, which controls the choice of the tokens that shall appear in the output. Despite their significance, the role of these parameters is often overlooked.

This paper systematically studies the impact of these parameters, as well as the number of prompt repetitions required to account for non-determinism,
in the context of 548 Java methods. We observe significantly different performances across different configurations of ChatGPT, with temperature having a marginal impact compared to the more prominent influence of the top-p parameter. Additionally, we show how creativity can enhance code generation tasks. Finally, we provide concrete recommendations for addressing the non-determinism of the model.

\end{abstract}


\section{Introduction} \label{sec:introduction}

LLMs are becoming increasingly adopted as assistants in code development tasks. 
For instance, the GitHub Copilot extension~\cite{Copilotvscode} for Visual Studio Code~\cite{vscode} has surpassed 22 million installations, making it 
one of the most widely used coding extensions overall in this IDE (Integrated Development Environment). Similarly, plugins for integrating LLMs, such as ChatGPT~\cite{ChatGPT} and Gemini~\cite{gemini}, are also increasingly becoming available for the most popular IDEs~\cite{codegpt,EclipseGemini}. 

Empirical studies demonstrate that AI assistants effectively generate implementations that assist developers in understanding, documenting, and validating their code. For instance, Corso et al.~\cite{Corso:EmpiricalAssessment:ICPC:2024} report that up to one-third of the method implementations can be generated automatically using LLMs. Similarly, the study by Pinto et al.~\cite{10.1145/3644815.3644949} reports positive feedback from the developers who use these tools. The study by Liang et al.~\cite{liang2024large} shows that, although the usability of these tools must still be improved, they are widely used with a positive impact on software development. Tufano et al.~\cite{10.1145/3643991.3644918} analyzed GitHub commits, pull requests, and issues showing how GitHub developers are using ChatGPT for several tasks, including feature implementation and enhancement, documentation, and testing.

The recommendations generated by LLMs are however dependent on several parameters that affect the \emph{correctness} and the \emph{determinism} of the responses. For instance, the \emph{temperature} can be used to control the creativity of the models, while the parameter $\textit{top-p}$ can be used to control the selection of the tokens that shall compose the output. Further, the non-determinism of the models requires multiple repetitions for each request to be assessed.

Despite the role played by these parameters, \emph{how the models are configured when used to generate code is underrated}. Studies often do not report the values used for model parameters, nor do they report the number of repetitions completed to assess the results~\cite{sakib2024extending,Siddiq:QualityCghatGPT:2024:MSR}. Sometimes studies try to minimize the non-determinism of the models to avoid doing many repetitions (e.g., setting the temperature to $0.0$), without considering the consequences of this decision on the correctness of the results~\cite{Yihong:CodeGenerationChatGPT:TOSEM:2024,Yue:RefiningChatGPTGeneratedCode:TOSEM:2024}. This spectrum of cases generates issues with the results themselves and their reproducibility. 

Early evidence confirms that these parameters and repetitions matter. For instance, the study by Arora et al.~\cite{arora2024optimizing} investigates the role of temperature and top-p when GPT is prompted to generate the body of a method from its signature. The study considers only 13 methods, reporting a significant role of both the temperature and top-p on the correctness of the responses. Moreover, low values of these parameters seem to reduce creativity, while increasing the consistency and correctness of the results. Similar results have been reported in Liu et al.~\cite{liu2024your}, where again low values of temperature and top-p are suggested to be beneficial for the correctness of the responses. The temperature has been specifically studied also in Ouyang et al.~\cite{ouyang2023empirical}, again confirming that low values may guarantee more consistent results from GPT models.  

Although these studies provide initial evidence of how these parameters may influence results in code generation tasks, they do not consider the \emph{impact of these parameters from the perspective of the non-deterministic nature of the models}, which imposes prompting the models multiple times to assess their effectiveness. For example, a low-temperature value may increase the consistency of the responses, not necessarily generating a proper implementation of a higher number of methods across repetitions. In particular, as discovered in this study, higher temperature values may \emph{decrease the absolute percentage of correct results} (i.e., we observe fewer correct responses to prompts, as reported in other studies), but may \emph{increase the range of methods correctly implemented} (i.e., creativity in responses enables the generation of implementations that can hardly be obtained otherwise). 

Practical \emph{recommendations on the number of repetitions} for each prompt are also lacking. Indeed researchers and practitioners make very different decisions, with little empirical justification. 

This paper addresses both these concerns by presenting an empirical study that systematically considers the impact of temperature and top-p in method generation tasks. The effect of these parameters is assessed in a context where the same inputs are submitted multiple times, and the range of responses generated is studied to derive practical implications about the \emph{setup and usage of these models}. We investigate these questions using ChatGPT, 
which is the most widely used LLM among developers \cite{MostUsedAI}, as well as one of the most frequently studied LLMs in various other research \cite{liu2024your,ouyang2023empirical,Yihong:CodeGenerationChatGPT:TOSEM:2024,Siddiq:QualityCghatGPT:2024:MSR,Yue:RefiningChatGPTGeneratedCode:TOSEM:2024,10.1145/3643991.3644918}.



More in detail we carefully selected a corpus of $548$ methods whose implementation is used to query GPT-4o, the most recent version of GPT at the time this paper has been written. Each query is submitted multiple times for multiple configurations, for a total of $27,400$ requests assessed. The analysis revealed surprising results that may influence future usage of LLMs in code generation. In particular:
\begin{itemize}
    \item Contrarily to current belief, using low-temperature values, although resulting in a higher number of requests responded correctly, decreases the total number of methods that could be implemented correctly considering repetitions (i.e., considering the submission of the same prompt multiple times).
    \item Although most of the studies report the value of the temperature used in the evaluation without mentioning the value of the top-p parameter, our study shows that top-p has a much stronger impact on the results than the temperature value, which has a marginal impact.
    \item  Submitting the same prompt multiple times is fundamental to accommodate for the non-deterministic nature of the queried models.
\end{itemize}

The study results in practical recommendations that may help researchers better set up and report their studies and may help practitioners use LLMs more effectively to generate code.

In a nutshell, our study provides the following contributions:
\begin{itemize}
\item Systematically examines how temperature and top-p affect method generation tasks in GPT-4o, releasing novel empirical evidence about the effect of these parameters on the correctness of the generated code.

\item Studies how to define the number of repetitions appropriately, to cost-effectively generate code.

\item Presents actionable recommendations for effectively configuring and utilizing GPT models in code generation tasks.

\item Releases an open dataset that can be used to reproduce the results reported in the paper and to conduct additional studies on the same subject.
\end{itemize}

The paper is organized as follows. Section~\ref{sec:methodology} presents the research questions we investigated and the methodology we designed to answer them. Section~\ref{sec:results} describes the results that we obtained to answer the research questions. Section~\ref{sec:related} discusses related work. Section~\ref{sec:conclusions} provides final remarks.

\section{Methodology} \label{sec:methodology}
In this section, we first discuss the parameters that can be defined by the user of ChatGPT and that may influence the results (Section~\ref{sec:parameters}). We then introduce the investigated research questions (Section~\ref{sec:rq}). We describe the dataset that we built to study the research questions (Section~\ref{ßec:dataset}). We finally report the experimental process that we defined to answer the research questions (Section~\ref{sec:process}).

\subsection{Parameters Influencing the Responses} \label{sec:parameters}

Some key parameters that influence the responses produced by GPT models are:
\begin{itemize}
\item \emph{Temperature} is often referred to as the parameter that determines the level of creativity of a model. In practice, it is a variable between $0.0$ and $2.0$ that affects the probabilities of occurrences of tokens, with values above $1.0$ inflating the probabilities of less likely tokens, and values below $1.0$ focusing on the most likely tokens only.

\item \emph{Top-p} is a parameter ranging between $0.0$ and $1.0$, controlling the number of tokens considered for the next token prediction\footnote{Not all implementations support $0.0$, some models may require a small but positive \emph{top-p} value.}. In particular, the model considers only enough tokens whose cumulative probabilities sum up to \emph{top-p} when predicting the next token~\cite{Holtzman:TopP:ICLR:2020}. Thus, the model considers fewer tokens for low \emph{top-p} values, and more tokens for high \emph{top-p} values. 

\item \emph{Frequency penalty} is a parameter ranging between $-2.0$ and $+2.0$, assigning a positive (negative values of the parameter) or a negative (positive values of the parameter) reward for repeating tokens either in the prompt or in the response.

\item \emph{Presence penalty} is the same as the frequency penalty but it applies to the response only.

\end{itemize}

Since in code generation tasks, the same tokens may occur multiple times (e.g., the same variables occurring multiple times in a method implementation), we do not introduce any penalty using a value of $0.0$ for frequency penalty and presence penalty, while we systematically study the impact of temperature and top-p.

\subsection{Research Questions} \label{sec:rq}

With this study, we want to answer four key research questions about the impact of temperature and top-p, as well as the number of responses that must be collected for the same prompts to account for the non-determinism of the model properly. We also analyze if any interaction exists between the parameter values and the method's complexity and length.

\textbf{RQ1 - What is the impact of \emph{Temperature} on the correctness of the results?} This research question considers multiple values of the temperature, namely $0.0$, $0.8$, $1.2$, and $2.0$, investigating its impact on the correctness of the results.

\textbf{RQ2 - What is the impact of \emph{Top-p} on the correctness of the results?} This research question considers multiple values of \emph{top-p}, namely $0.0$, $0.5$, and $0.95$, investigating its impact on the correctness of the results.

\textbf{RQ3 - How many times should the same prompt be submitted to likely collect the correct results?} This research question studies how the non-determinism of the models impacts the correctness of the results, identifying an optimal number of repetitions to be completed.

\textbf{RQ4 - How does the effectiveness of the studied configurations change with the complexity and length of the methods?} This research question studies if the configurations obtained with different values of temperature and top-p perform differently for methods of different cyclomatic complexity and length.

\begin{figure*}
\centering
\includegraphics[width=\textwidth]{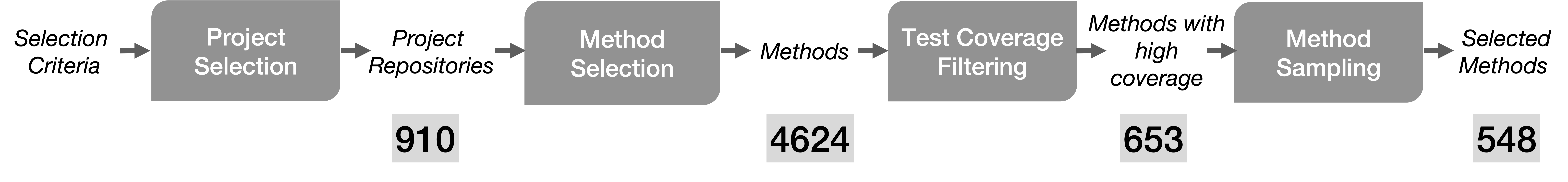}
\caption{Dataset construction process.}
\label{fig:methodology}
\end{figure*}

\subsection{Dataset Construction} \label{ßec:dataset}

To build a dataset of methods whose implementation has to be generated by ChatGPT, we analyzed GitHub looking for relevant repositories and methods. The process applied to construct the dataset is illustrated in Figure~\ref{fig:methodology}.

We started by using the GitHub API to select all the repositories of Java applications ranked with at least 500 stars, to guarantee the popularity of the repository, and with commits occurring in the time window from October 3rd, 2023 to June 6th, 2024. Since GPT-4o has been trained before October 2023, this time window prevents any data leakage. To ease the automatic analysis of the code, we selected maven projects only. These criteria returned 910 eligible repositories.  

From this set of repositories, we looked for commits created between October 3rd, 2023 and June 6th, 2024 that add a full method implementation (i.e., the method has been added to the repository after GPT-4o has been trained) of at least 5 lines (to prevent dealing with trivial methods) in \texttt{\small src/main} (to focus on the core code of applications, discarding tests and auxiliary methods). Since we use the Javadoc comment as a method specification in the prompt created for ChatGPT, we only select the methods with at least five lines of Javadoc. Finally, since we need test cases to check the correctness of any plausible implementation of the method, we only select a method if there exists a test class with test cases in \texttt{\small src/test} associated with the selected method. After applying these criteria, we obtained 4,624 methods.

We check out all the selected methods and associated tests and compute the percentage of statements in the methods covered by the tests using JaCoCo\footnote{\url{https://www.eclemma.org/jacoco/}}. We discard methods whose coverage is below 80\% since the tests are supposed to not exercise well enough the behavior of the method, and thus they would not be a reliable proxy of correctness for any candidate implementation. From this step, we obtained 653 methods, whose coverage ranges from 80\% to 100\% with an average value of 97.4\%.

To make the final selection of the methods to be used for the study, we compute their cyclomatic complexity using Understand\footnote{\url{http://scitools.com}}. We then divide the methods into buckets, one bucket for each complexity value. Inside each bucket, we ordered the available methods first by the number of stars of the repository they have been extracted from, then by coverage value (giving higher priority to the methods with higher coverage), and finally by the commit date (higher priority to the most recent methods). We finally selected a total of 548 methods that are well distributed in terms of complexity values for our study. We selected a different number of methods from each bucket to reflect the non-uniform size of the buckets: method complexity ranges from 1 to 38, and as complexity increases, the number of methods in the bucket decreases.

\subsection{Experimental Process} \label{sec:process}

\emph{To answer RQ1}, for each method in the dataset, we submit to ChatGPT the request to produce an implementation starting from its signature, Javadoc descriptions, and the rest of the code in the class as context.
We repeat this request ten times, to account for non-determinism. Moreover, we repeat this process four times using four temperature values: $0.0$, $0.8$ (default value at the time we performed the study), $1.2$, and $2$. We keep top-p to its default value of 0.95. This results in a total of 27,450 method implementations collected (i.e., 4 temperature values $\times$ 10 repetitions $\times$ 548 methods).

To assess the quality of the recommended code, we distinguish three cases that can be determined automatically:

\begin{itemize}

\item \emph{Invalid}: the returned code does not compile.

\item \emph{Incorrect}: the returned code compiles but fails at least one of the available test cases.


\item \emph{Plausible}: the returned code compiles and passes all the available test cases. 
\end{itemize}

We label the code as plausible and not correct since passing all the available test cases does not guarantee the correctness of the resulting code. On the other hand, the fact that the returned code compiles and passes a set of test cases exercising at least 80\% of the code in the implementation (97.4\% on average) provided by the developers represents a good indicator of the quality of the returned code. We relied on this proxy measure of correctness to complete a large-scale assessment of ChatGPT.

To determine the role of the parameter temperature, we compare the rate of invalid, incorrect, and plausible implementations returned while changing the temperature.  

\emph{To answer RQ2}, we select the temperature value producing the most promising results with RQ1 and submit the same prompts to ChatGPT using different top-p values: $0.95$ (already used in RQ1), $0.5$, and $0.0$. We execute each prompt five times to account for non-determinism. This leads to a total of 4,480 additional method implementations collected (i.e., 2 more top-p values $\times$ 5 repetitions $\times$ 548 methods).

Again, we compare the results considering the rate of invalid, incorrect, and plausible implementations returned while changing top-p.  

\emph{To answer RQ3}, we analyze the data collected for RQ1 and RQ2, considering the number of repetitions needed to collect a correct answer. In particular, we compute the $pass@k$ metric that returns, for a given method, the probability of collecting at least a plausible implementation with $k$ submissions of the same prompts. 

For each method, we compute the value $pass@k$, for $k=1\ldots n$, where $n=10$ is the number of repetitions we completed. In particular, we  compute the metric as recommended by Lyu et al.~\cite{lyu2024top} according to the formula

\[
\text{\textit{pass@k(m)}} = 1 - \dfrac{\dbinom{n - c}{k}}{\dbinom{n}{k}}
\]

\noindent
where $c$ is the number of plausible method implementation for method $m$, and $n=10$ is the total number of repetitions. Note that \(\binom{n - c}{k}\) represents the number of combinations of $k$ sequential responses without any plausible implementation, and \(\binom{n}{k}\) is the total number of $k$ sequential responses that can be returned. 

Based on the value of the metric $pass@k$, we discuss the values of $k$ that could guarantee a good compromise between the number of repetitions and the probability of collecting a plausible implementation.

\emph{To answer RQ4}, we analyze the data collected for RQ1 and RQ2, studying the percentage of plausible answers provided for methods of increasing cyclomatic complexity and length. We also compute the Pearson correlation coefficient, to determine the existence of any correlation between the rate of plausible method implementations generated, and the complexity and length of the methods. The analysis is repeated for each studied configuration.

\subsection{Tool Support}
We automated both the dataset construction process and the prompt submission process, to scale to the magnitude required by this study. Both tools are implemented as a set of Python scripts that we released publicly, jointly with our dataset, the response generated by ChatGPT, and the code to generate the visualizations used in this paper. The experimental material is available in the following repository \urlRepo.

\section{Results} \label{sec:results}

\subsection{RQ1 - Temperature}

\begin{figure}[ht]
        \centering
        \includegraphics[width=0.5\textwidth]{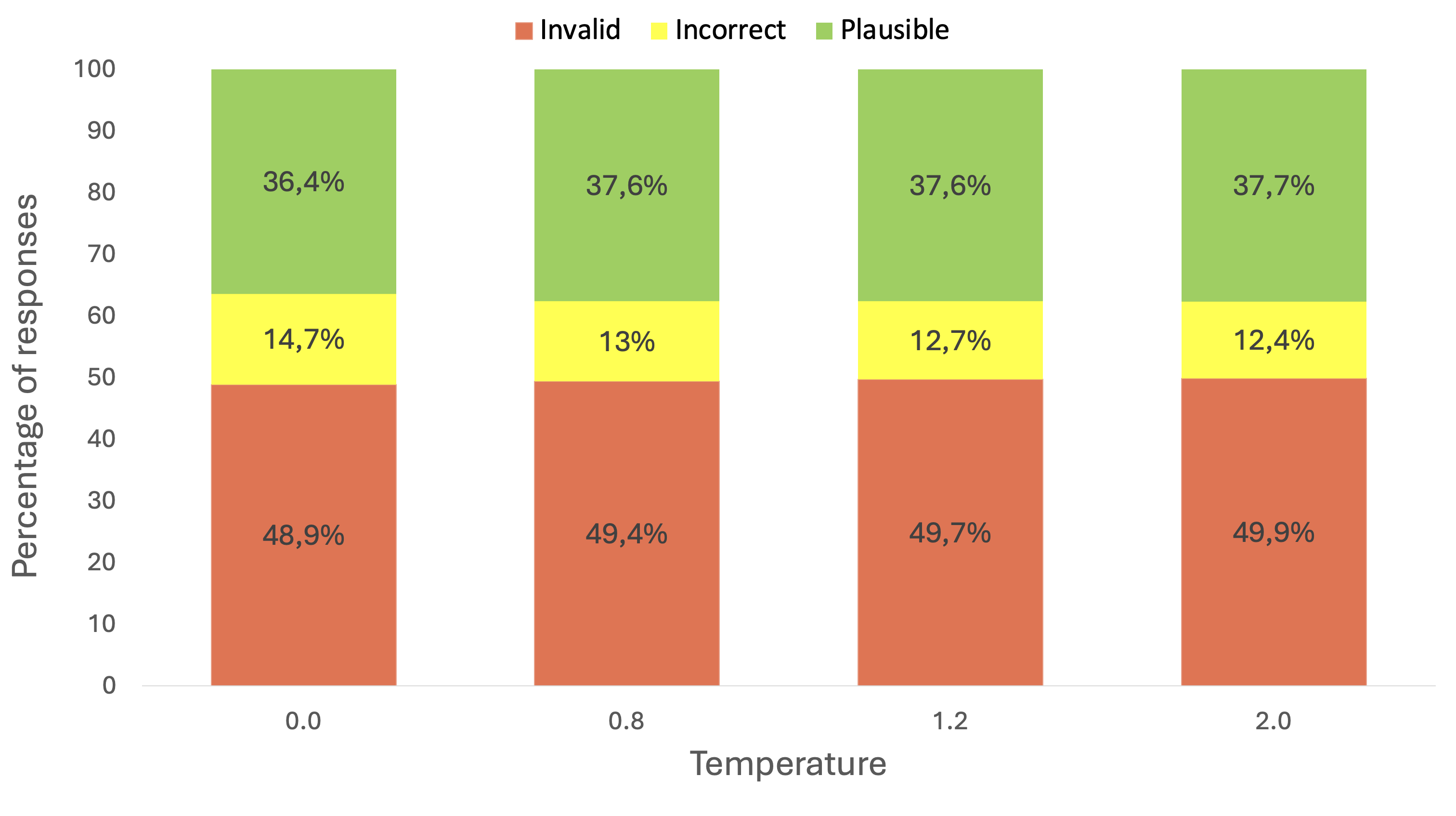}
        \caption{Percentage of invalid, incorrect, and plausible responses for different \emph{temperature} values.}
        \label{fig:RequestsCorrectness}
\end{figure}

Figure~\ref{fig:RequestsCorrectness} shows the percentage of invalid, incorrect, and plausible responses generated with different temperature values. ChatGPT tends more to generate either invalid or plausible code, with a smaller percentage (between 12\% and 14.7\%) of incorrect code, that is, if the code compiles there is a good change it also passes the test cases. Response-wise, differences are small and non-significant, with a percentage of valid implementations ranging between 36.4\% (temperature $0.0$) and 37.7\% (temperature $2.0$).

\begin{figure}[ht]
        \centering
        \includegraphics[width=0.5\textwidth]{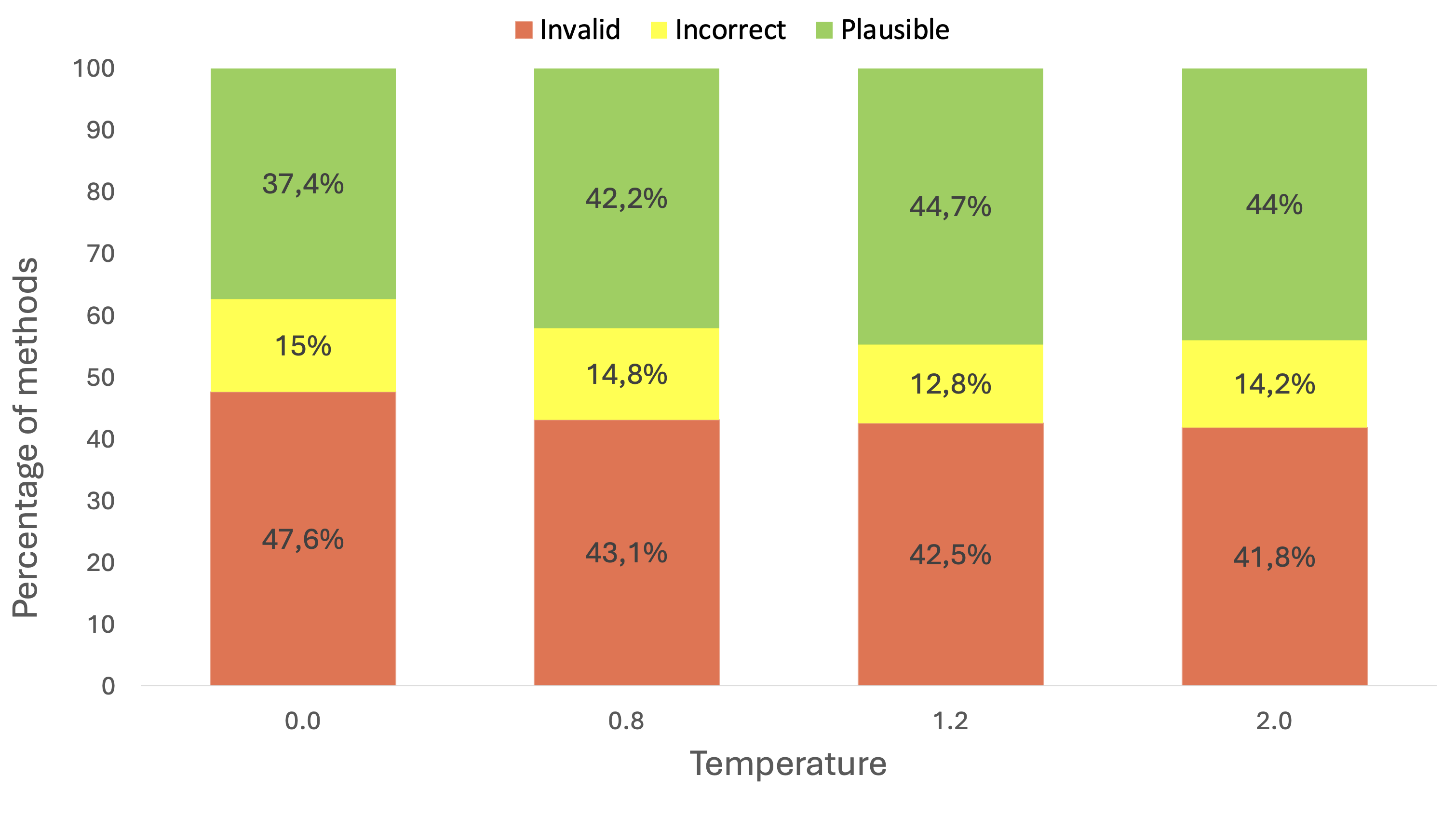}
        \caption{Percentage of \emph{methods} with at least a plausible, incorrect, or invalid implementation, considering the best result returned by a configuration across 10 repetitions.}
        \label{fig:MethodsCorrectness}
\end{figure}

However, if we consider the capability of the multiple temperature values to generate a plausible implementation at least once in the scope of the ten repetitions of the same request, the model's performance changes for different temperature values. In particular, Figure~\ref{fig:MethodsCorrectness} shows for each temperature value the percentage of methods with at least a plausible implementation in the ten repetitions.  It is apparent how higher temperature values can generate a correct implementation for a higher number of methods, with temperatures $1.2$ and $2.0$ achieving the best results.

Considering the almost invariant number of responses with plausible method implementation (Figure~\ref{fig:RequestsCorrectness}) paired with the higher number of methods with at least one plausible implementation generated (Figure~\ref{fig:MethodsCorrectness}), we can conclude that the higher creativity obtained with higher temperature values lets ChatGPT generates plausible code for methods that cannot be otherwise generated, at the cost of decreasing the rate of plausible code for the easier methods.  

\begin{figure}[ht]
        \centering
        \includegraphics[width=0.5\textwidth]{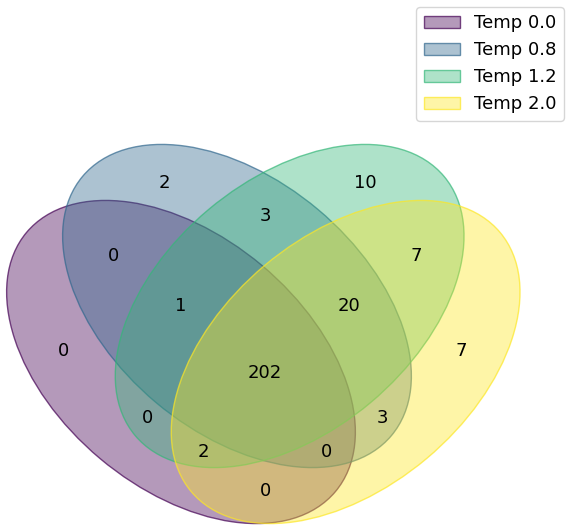}
        \caption{Eulero-Venn diagram with the plausible methods generated at least once for each temperature value.}
        \label{fig:EuleroVenn}
\end{figure}

This is confirmed by the Eulero-Venn diagram of the methods with at least one plausible implementation for each temperature value shown in Figure~\ref{fig:EuleroVenn}, where temperatures $1.2$ and $2.0$ generated plausible code at least once for the highest number of methods, nearly subsuming the results obtained for temperatures $0.0$ and $0.8$.  

Indeed, by thoroughly sampling the generated code, it is possible to recognize how using low-temperature values may cause ChatGPT to systematically fail the generation of code, even for tasks that might initially appear to be relatively straightforward. For instance, the code displayed in Listing~\ref{lst:example1} serves as an example of a method that was consistently generated incorrectly across all ten repetitions, with ChatGPT repeatedly making a subset of the same mistakes in each attempt. Specifically, ChatGPT often failed to include the main parentheses of the method (lines 1 and 7), incorrectly reused variable names by changing \texttt{columnCount} to \texttt{column} (lines 2 and 4), used parentheses without providing arguments (line 2), and frequently assigned the wrong name to the method.
Conversely, a plausible and correct implementation of this method becomes achievable when higher temperature values are used (e.g., temperatures of $1.2$ and $2.0$). Under these configurations, the model's enhanced creativity is effectively leveraged, allowing it to overcome its inherent bias toward generating flawed implementations.

\begin{lstlisting}[language=Java,
xleftmargin=.12in,
label=lst:example1,caption=Invalid code obtained with Temperature 0.0.]
private static boolean isJaggedMatrix(double[][] m) 
    int columnCount = m[].length;
    for (double row : m) {
        if (row.length != column) {
            return
        }
    }
\end{lstlisting}

\textbf{\textit{In a nutshell}}, although creativity has a relatively marginal impact on the correctness of individual responses on a per-request basis, it has proven to be valuable when considering the overall capability of generating plausible code for each method. Specifically, a temperature setting of $1.2$ yielded slightly better results compared to $2.0$ in terms of the total number of plausible methods successfully addressed.

\subsection{RQ2 - Top-p} 

\begin{figure}[ht]
        \centering
        \includegraphics[width=0.5\textwidth]{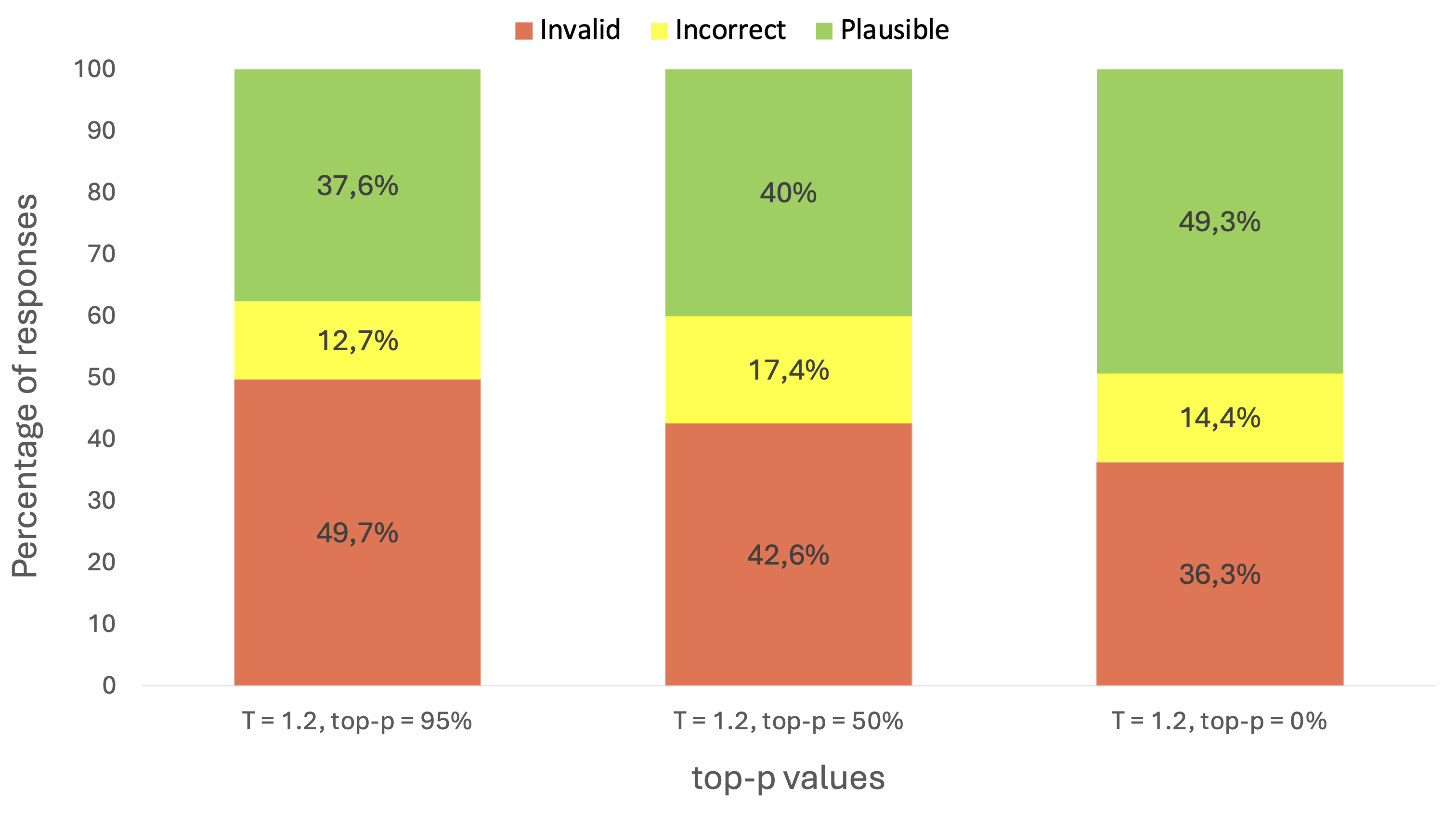}
        \caption{Percentage of invalid, incorrect, and plausible responses for different values of \textit{top-p}.}
        \label{fig:RequestsTopP}
\end{figure}

Figure~\ref{fig:RequestsTopP} illustrates the percentage of invalid, incorrect, and plausible code returned by the requests when using temperature equal to $1.2$, which was the best-performing temperature value in RQ1, and multiple values of the top-p parameter. The results indicate that the top-p parameter exerts a considerably more significant influence on the generation of plausible code compared to the temperature setting. 

On one hand, the proportion of invalid code exhibits a noticeable decrease, moving from 49.7\% when top-p is set to $0.95$, down to 42.6\% and 36.3\% for top-p equal to $0.5$ and $0.0$, respectively. On the other hand, the percentage of plausible code generated increases correspondingly, rising from 37.6\% for top-p equal to $0.95$, to 40.0\% and 49.3\% for top-p equal to $0.5$ and $0.0$, respectively.

\begin{figure}[ht]
        \centering
        \includegraphics[width=0.5\textwidth]{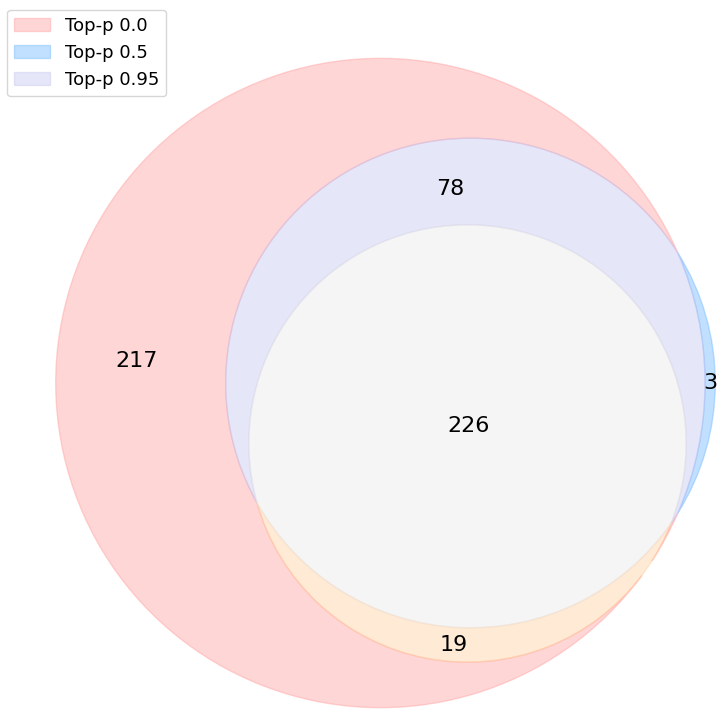}
        \caption{Eulero-Venn diagrams of the plausible methods generated at least once with each \textit{top-p} value.}
        \label{fig:MethodsTopP}
\end{figure}

The dramatic positive impact of low top-p values is clearly confirmed when we focus on the methods that have a plausible implementation in at least one of the repetitions.  Figure~\ref{fig:MethodsTopP} illustrates the Eulero-Venn diagram representing the number of different plausible methods that were generated at least once by each specific top-p value configuration. 

The smaller the top-p value, the more effective the configuration appears to be in practice. In particular, when using top-p set to $0.0$, there are $217$ plausible methods that are uniquely generated by this configuration, which also subsumes the other studied configurations, with the exception of $3$ plausible methods generated with top-p equal to $0.5$. This outcome aligns well with previous studies, which have consistently reported a higher effectiveness associated with low top-p values~\cite{arora2024optimizing}.

\begin{lstlisting}[language=Java, xleftmargin=.12in,label=lst:example2,caption= Systematically wrong implementation of the \texttt{evaluate}\\ method with \textit{top-p} 0.95.] 

public boolean evaluate(Geometry b, 
                    TopologyPredicate predicate) {
    if (b.isEmpty()) {
        return true;
    }
    return false;
}
\end{lstlisting}

Indeed, by sampling the generated code, it is possible to recognize how using high top-p values may cause ChatGPT to systematically fail the generation of code. For example, with top-p equal to $0.95$, ChatGPT never 
adds some of the fundamental checks required in the function 
\texttt{\small public boolean evaluate(Geometry b, TopologyPredicate predicate)}, thereby producing an incomplete or incorrect implementation for each attempt. 
Listing~\ref{lst:example2} shows a typical faulty implementation where conditions on \texttt{Geometry b} and \texttt{TopologyPredicate predicate} are not properly handled or are entirely omitted.

On the contrary, when using top-p equal to $0.0$, ChatGPT tends to be more consistent and systematically covers the necessary logic for the same method. In Listing~\ref{lst:example3}, we show a plausible implementation that checks whether \texttt{b} is valid and non-empty, and handles multiple values of \texttt{predicate} through proper conditional structures.

\begin{lstlisting}[language=Java, xleftmargin=.18in, label=lst:example3, caption=Plausible implementation with \textit{top-p} 0.0.]
public boolean evaluate(Geometry b, 
                    TopologyPredicate predicate){
  if (b == null || b.isEmpty()) 
    return false;
  switch (predicate) {
    case DISJOINT: 
        return !b.intersects(this.geometry);
    case TOUCHES:  
        // ...
    default:       
        return false;
  }
}
\end{lstlisting}

\textbf{\textit{In a nutshell}}, we observe top-p playing a bigger role in method generation than temperature, with smaller top-p values being more effective than higher values. 
In addition to setting low top-p values, these findings highlight the importance of reporting the full configuration of ChatGPT, not just the temperature setting, as results may vary significantly depending on the set of tokens considered for generating the response, which is controlled by the top-p parameter.

\subsection{RQ3 - Repetitions}
\begin{figure}[ht]
        \centering
        \includegraphics[width=0.5\textwidth]{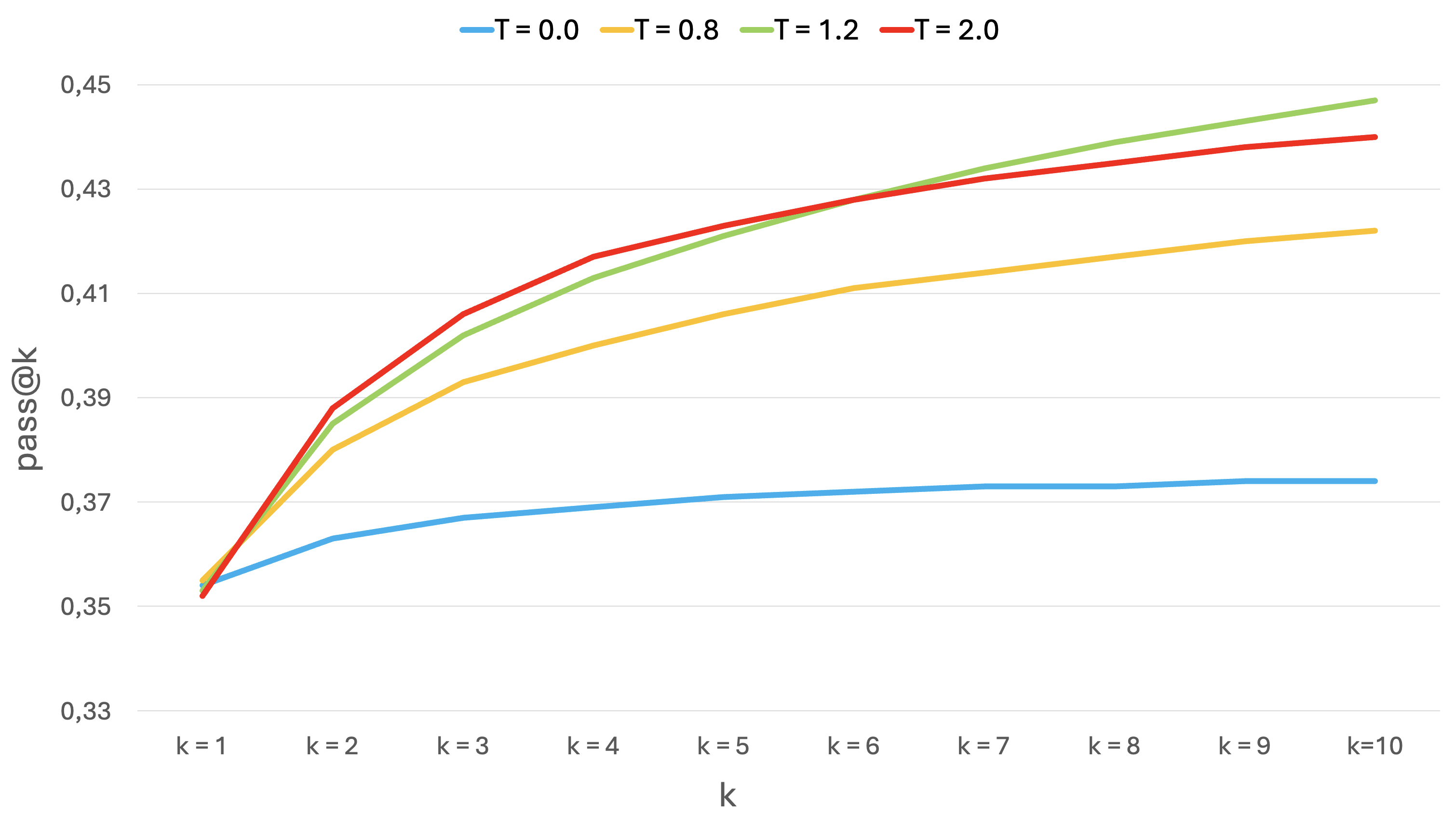}
        \caption{Value of pass@k for multiple temperature values.}
        \label{fig:passKTemperature}
\end{figure}

With this research question, we investigate the probability of observing a plausible solution in the scope of multiple repetitions, that is, the multiple submissions of the same prompt. Figure~\ref{fig:passKTemperature} presents the pass@k metric, displaying its values for increasing values of $k$ and across different temperature configurations.

While there is almost no difference in the performance of ChatGPT for individual requests, we can see how the performance of the model changes when multiple repetitions are considered. In particular, temperature $0.0$ leads to the discovery of little additional plausible implementations when increasing the number of repetitions. On the contrary, the other studied temperature values can discover additional plausible implementations for increasing values of $k$. 

This result suggests that, although using a temperature equal to $0.0$ may motivate empirical setup with no, or very little, repetitions, it also affects the effectiveness of the result.  

We can observe how higher creativity is useful to cover a broader range of implementations, with a value equal to $1.2$ representing a good compromise between creativity and the soundness of the behavior of the model. 

\begin{figure}[ht]
        \centering
        \includegraphics[width=0.5\textwidth]{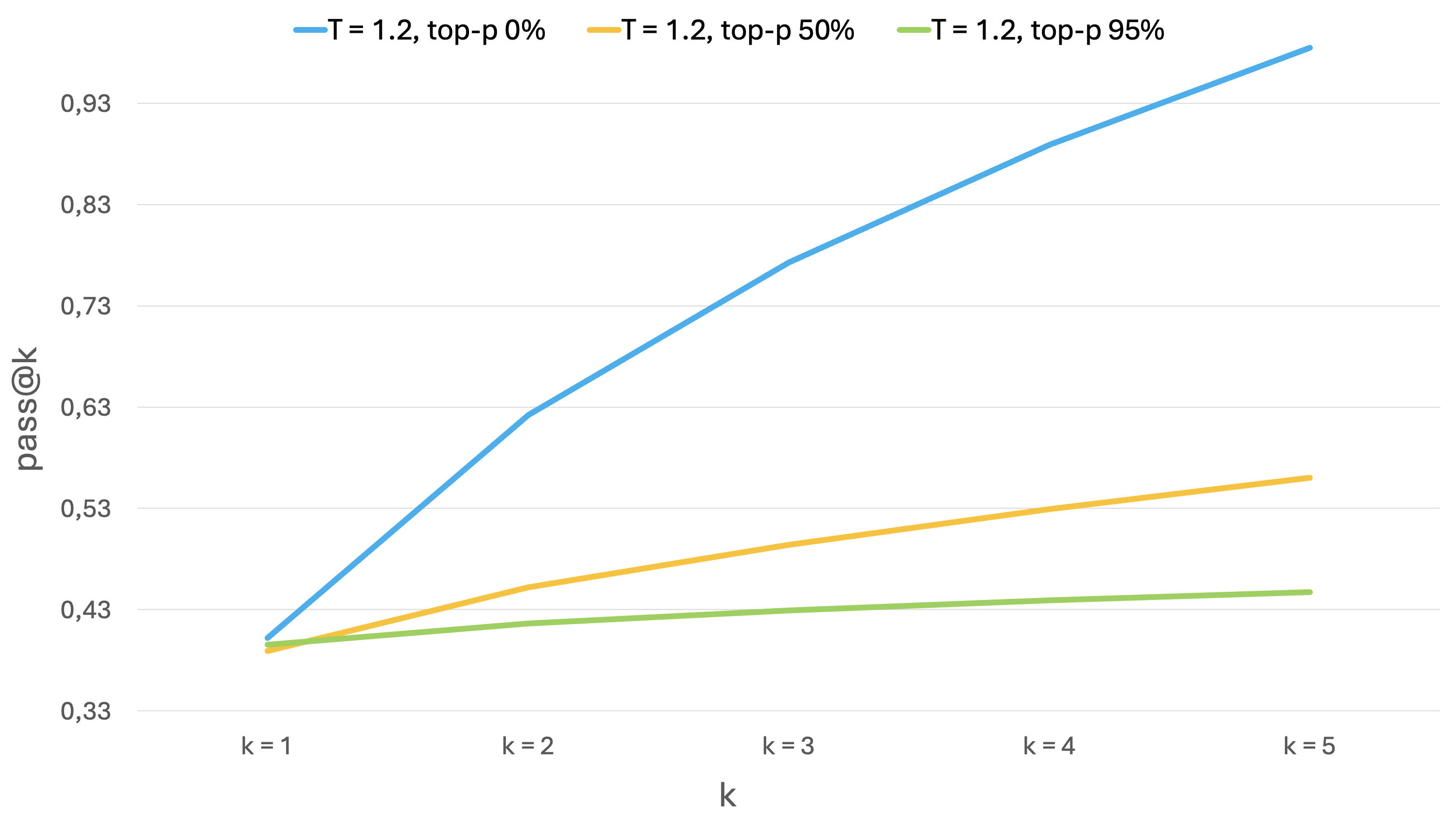}
        \caption{Value of pass@k for multiple \textit{top-p} values.}
        \label{fig:passKTopP}
\end{figure}

We also studied the combination of repetitions and top-p. Figure~\ref{fig:passKTopP} shows how values of pass@k change with an increasing number of repetitions.

The results confirm the relevance of the top-p parameter and its impact on the capability of generating plausible code. In particular, while top-p values equal to $0.95$ and $0.5$ saturate quite quickly, a top-p value of $0.0$ allows the model to generate plausible code for the vast majority of the methods. Notably, five repetitions yield the best results. 

To illustrate why multiple attempts can be critical, let us consider the \texttt{checkSymmetry} method, which verifies if a given matrix is symmetric. In the first attempts, ChatGPT produces an incomplete or incorrect implementation, as shown in Listing~\ref{lst:example4}, where the code checks only the first and last elements of each row instead of comparing \texttt{matrix[i][j]} and \texttt{matrix[j][i]}:

\begin{lstlisting}[language=Java, xleftmargin=.12in, label=lst:example4, 
caption=Example of an incorrect implementation of\\ \texttt{checkSymmetry} in the early attempts.]
public boolean checkSymmetry(int[][] matrix) {
    for (int i = 0; i < matrix.length; i++) {
        if (matrix[i][0] != matrix[i]
                     [matrix.length - 1]) {
            return false;
        }
    }
    return true;
}
\end{lstlisting}

By repeatedly submitting the same prompt (e.g., up to the fifth attempt), the model finally produces a plausible version of the method. Listing~\ref{lst:example5} shows how the code correctly ensures that the matrix is square and systematically checks all symmetrical pairs of elements:
\begin{lstlisting}[language=Java, xleftmargin=.18in, label=lst:example5, 
caption=Plausible implementation of \texttt{checkSymmetry} obtained after\\ five attempts.]
public boolean checkSymmetry(int[][] matrix) {
    if (matrix == null || matrix.length == 0) {
        return false;
    }
    int n = matrix.length;
    for (int i = 0; i < n; i++) {
        if (matrix[i].length != n) {
            return false;
        }
        for (int j = i + 1; j < n; j++) {
            if (matrix[i][j] != matrix[j][i]) {
                return false;
            }
        }
    }
    return true;
}
\end{lstlisting}

These two snippets highlight how a single attempt may fail, but repeated submissions can yield a functional implementation that passes all test cases, thereby improving the pass@k performance.

\textbf{\textit{In a nutshell}}, it is important to submit the same prompt multiple times to address the non-determinism of the model. Low-temperature values reduce the need to run multiple repetitions, yet require submitting a prompt more than once. However, low-temperature values also decrease the range of plausible implementations that can be obtained. The best trade-off is achieved by using a temperature equal to $1.2$, with top-p equal to $0.0$, and completing five repetitions.

\subsection{RQ4 - Impact of Methods Length and Complexity}

\begin{figure}[ht]
        \centering
        \includegraphics[width=0.5\textwidth]{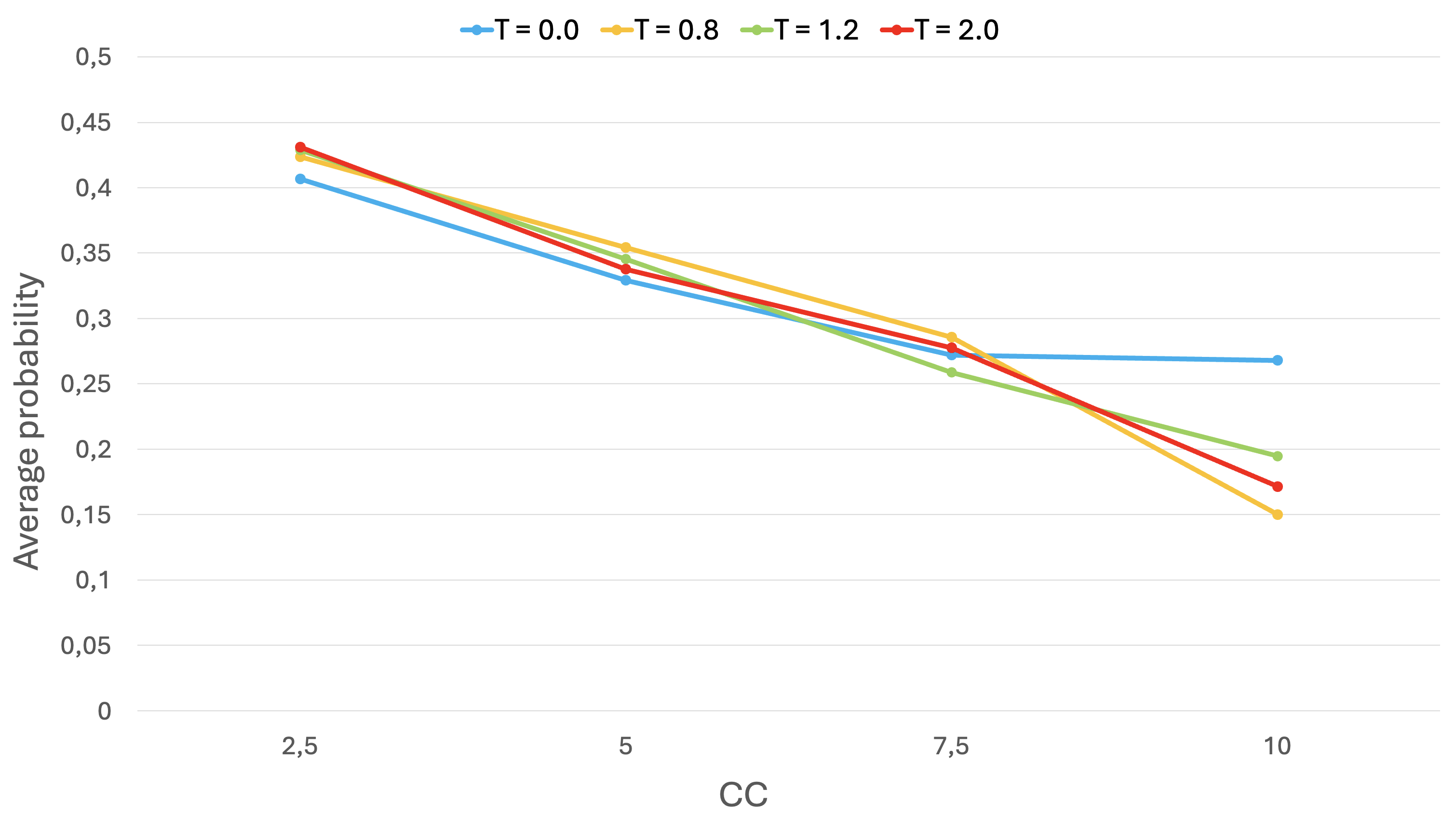}
        \caption{Average probability of producing a plausible implementation for increasing values of complexity.}
        \label{fig:AvgProbPlauvsCC}
\end{figure}

\begin{figure}[ht]
        \centering
        \includegraphics[width=0.5\textwidth]{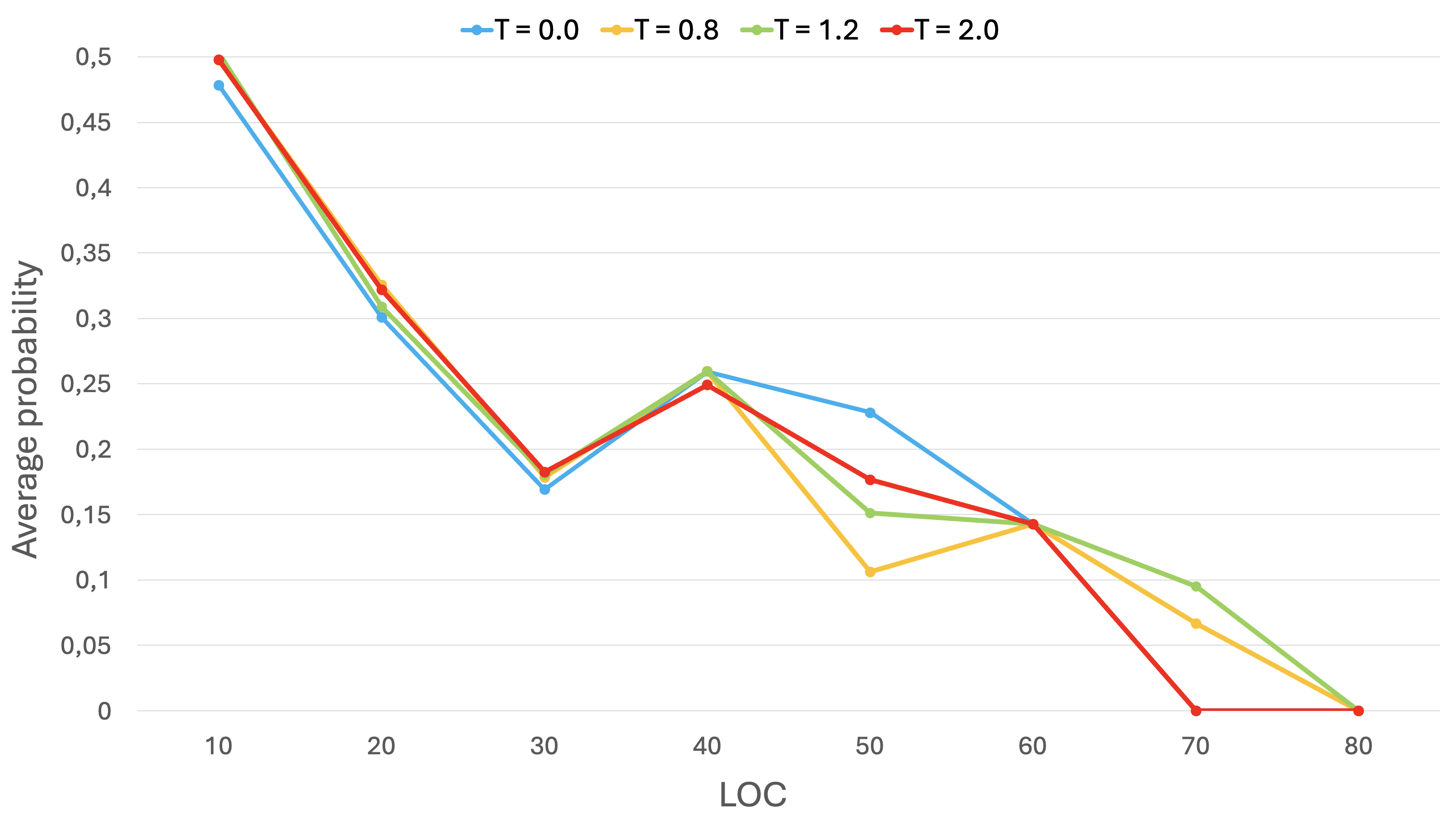}
        \caption{Average probability of producing a plausible implementation for increasing lengths.}
        \label{fig:AvgProbPlauvsLength}
\end{figure}

Figures~\ref{fig:AvgProbPlauvsCC}  and~\ref{fig:AvgProbPlauvsLength}
show how the average probability of producing a plausible response relates to the cyclomatic complexity and length of the method that has to be generated. For the cyclomatic complexity, the plot reports average probabilities using an interval of 2.5 points. For instance, the data point associated with 5 indicates the average probability of generating plausible code for the methods whose complexity is between 2.5 and 5. We report values only for complexity intervals that include at least 50 data points. For the length of the method, the plot reports exactly the same information, using an interval of 10 locs.

\begin{figure}[ht]
        \centering
        \includegraphics[width=0.48\textwidth]{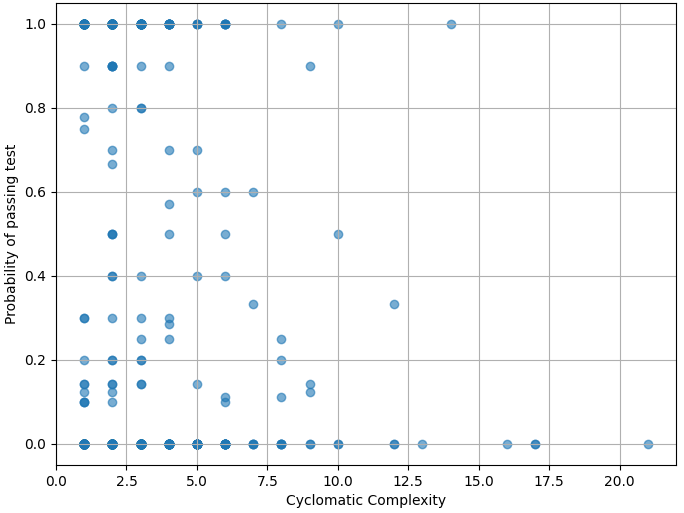}
        \caption{Scatter plot of probabilities for temperature 1.2.}
        \label{fig:ScatterProbT1.2}
\end{figure}

We can notice how the trend is strongly linear. However, the linear trend of average probabilities is not linear when considering the individual probabilities of producing plausible code for the methods in the dataset. In fact, Figure~\ref{fig:ScatterProbT1.2} shows the scatter plot of these probabilities for temperature 1.2. Many data points overlap. We can notice how the probabilities associated with the individual methods are very diverse, and the linear trend shows up only at the level of the overall average values, and it is not evident when considering the individual data points. All the plots are similar, also considering other temperature values and using method length rather than complexity. All the plots are available as part of our online material.

We checked for the presence of any correlation by computing the Pearson correlation coefficient and its significance. We report the results in Table~\ref{tab:correlation}.

\begin{table}[ht]
\caption{Pearson correlation coefficient and its significance for multiple temperature values in relation to cyclomatic complexity and lines of code. Significant p-values in bold.}

\centering
\begin{tabular}{ccc}

\textbf{Temperature} & \textbf{Corr. Coef.} & \textbf{P-Value}  \\
\toprule
\multicolumn{3}{c}{Correlation with CC} \\
\midrule
0.0 & -0.10 & \textbf{0.02} \\
0.8 & -0.12 & \textbf{0.01} \\
1.2 & -0.12 & \textbf{0.01} \\
2.0 & -0.11 & \textbf{0.01} \\
\toprule
\multicolumn{3}{c}{Correlation with locs} \\
\midrule
0.0 & -0.18 & \textbf{$<$0.01} \\
0.8 & -0.20 & \textbf{$<$0.01} \\
1.2 & -0.20 & \textbf{$<$0.01} \\
2.0 & -0.21 & \textbf{$<$0.01} \\
\bottomrule

\end{tabular}
\label{tab:correlation}
\end{table}

We can notice how the correlation is always significant, but its strength is low, with a slightly stronger correlation with method length. This is a direct consequence of the average linear trend combined with the high variability of the values. That is, this result indicates that the longer or the more complex methods are, the more difficult is to generate a plausible implementation. However, this observation is very sensitive to the individual cases, because there are lengthy and complex methods that are easy to generate (e.g., methods that resemble well-known algorithms), and short methods that are hard to generate (e.g., methods that use APIs that were not existing at the time of the training), making method length and method complexity inaccurate predictors of the chance of obtaining plausible code. 

We can finally observe how all the temperature values perform very similarly, indicating no major difference in the capability of dealing with methods of varying length and complexity.

\begin{figure}[ht]
        \centering
        \includegraphics[width=0.5\textwidth]{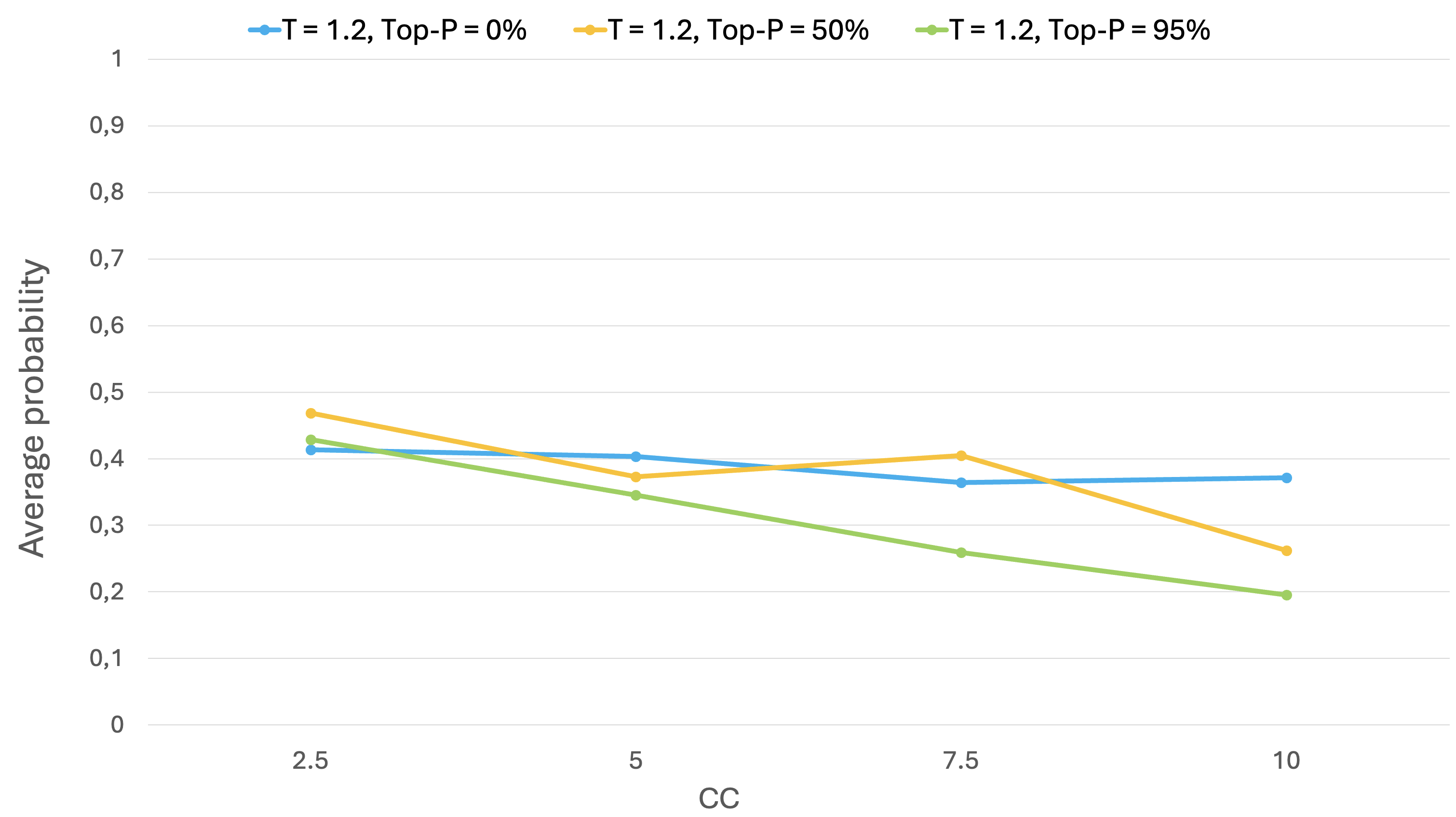}
        \caption{Average probability of producing a plausible implementation for increasing complexity.}
        \label{fig:corrCCTopp}
\end{figure}

\begin{figure}[ht]
        \centering
        \includegraphics[width=0.5\textwidth]{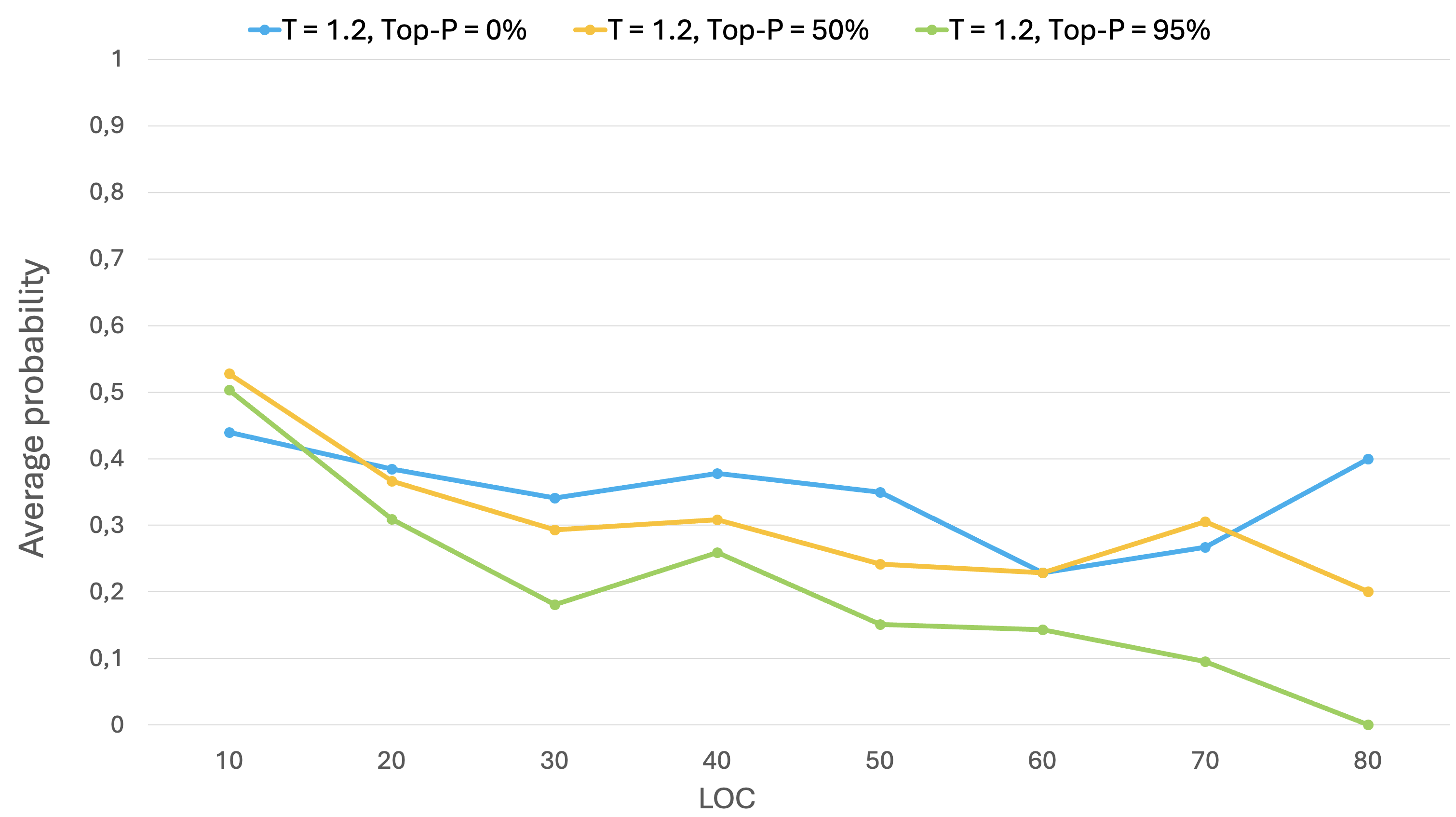}
        \caption{Average probability of producing a plausible implementation for increasing lengths.}
        \label{fig:corrLocTopp}
\end{figure}

We repeated the same analysis for top-p values. Figures~\ref{fig:corrCCTopp}  and~\ref{fig:corrLocTopp} show how the average probability of producing a plausible response relates to the cyclomatic complexity and length of the method that has to be generated, for different top-p values. The trend here is less evident, with lines that are quite flat. When considering individual probabilities, again they are very diverse spanning the full range of possible values (see for instance the results for top-p equal to $0.0$ in Figure~\ref{fig:corrCCTopp}).

\begin{figure}[ht]
        \centering
        \includegraphics[width=0.48\textwidth]{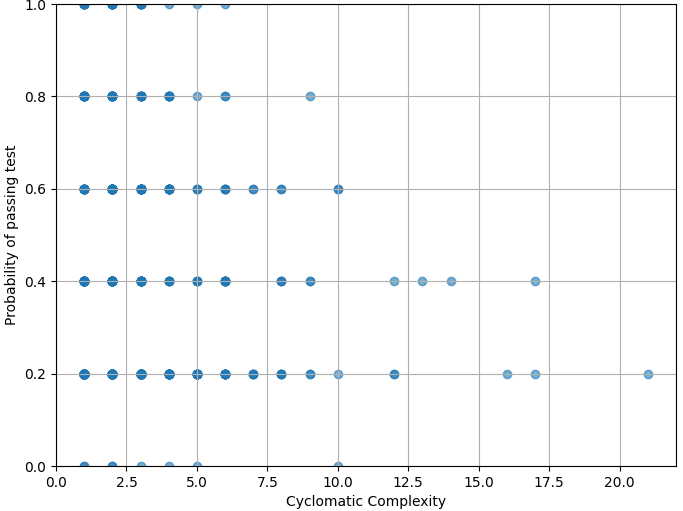}
        \caption{Average probability of producing a plausible implementation for increasing complexity.}
        \label{fig:corrCCTopp}
\end{figure}

We checked for the presence of any correlation by computing the Pearson correlation coefficient and its significance. We report the results in Table~\ref{tab:correlationTopP}.

\begin{table}[ht]
\caption{Pearson correlation coefficient and its significance for multiple temperature values in relation to cyclomatic complexity and lines of code. Significant p-values in bold.}

\centering
\begin{tabular}{ccc}
\textbf{Top-p} & \textbf{Corr. Coef.} & \textbf{P-Value}  \\
\toprule
\multicolumn{3}{c}{Correlation with CC} \\
\midrule
0.0 & -0.06 & 0.13 \\
0.5 & -0.09 & 0.08 \\
0.95 & -0.12 & \textbf{$<$0.01} \\
\toprule
\multicolumn{3}{c}{Correlation with locs} \\
\midrule
0.0 & -0.13 & \textbf{$<$0.01}  \\
0.5 &  -0.17 & \textbf{$<$0.01}  \\
0.95 & -0.20 & \textbf{$<$0.01}  \\
\bottomrule

\end{tabular}
\label{tab:correlationTopP}
\end{table}

We can observe how the size of the method correlates with the probability of generating plausible code, while the correlation with the cyclomatic complexity holds only for high top-p values. In practice, low top-p values increase the effectiveness of code generation, eliminating the (mild) correlation between the generation of plausible code with the method complexity. Interestingly, method length remains a relevant factor even for low top-p values. Lastly, the correlation is in all the cases mild, again confirming that there is a heterogeneity of cases and that both size and complexity are factors that cannot explain alone the effectiveness of code generation.

This evidence further confirms the relevance of the configuration in the code generation process and the importance of using low top-p to address a wider range of cases.

\textbf{\textit{In a nutshell}}, method length and complexity have a mild impact on the capability of generating plausible code. Contrarily to other configurations, low top-p values do not show any correlation with method complexity, only retaining the correlation with method length, confirming their better suitability for code generation tasks.




\subsection{Threats to Validity}
Our study faces several threats. One threat is the data leakage problem, that is, the model might have observed during the training phase the same methods we used in the evaluation. To address this threat, we carefully selected only the methods that have been committed on GitHub after the official model training date. We expect this practice to nearly eliminate the threat. The only possibility for the model to know the implementation of a method is the case of cross-repository code clones, that is, developers who cloned code produced before the training date from another repository. Although this is in principle possible, we expect it to occur rarely.

Another threat is about checking the responses produced by ChatGPT. The scale of our study, consisting of 548 methods whose implementations have been requested multiple times with different configurations, for a total of 27,400 implementations assessed, does not allow for the manual inspection of the code by one or more developers. For this reason, we relied on the notion of plausible method implementation, that is, an implementation that passes the test cases implemented by the developer to test the method, where the test cases are required to cover at least 80\% of the code (97.4\% in average) present in the original method implementation. We assume this is a notion strong enough to assess the relative performance of the various configurations and the impact of the parameters. We do not claim that plausible code is correct code, as shown in other studies there is a gap between plausible and correct code~\cite{Corso:EmpiricalAssessment:ICPC:2024}, but rather that plausible code is a useful metric to study the impact of the parameters on the responses.   

A final threat is the generality of our findings. Our study considers ChatGPT and the generation of full-method implementations. We do not know if the results are also valid for other assistants and tasks. However, we expect the results to remain valid as long as the assistants are based on the same model architecture and the task is the same. Additionally, our study targets Java code. This limitation may restrict the generalizability of our findings to other programming languages. In fact LLMs may exhibit different behaviors with different languages. 
%

\subsection{Findings}

Our study generates some actionable findings that are discussed below:

\begin{itemize}
\item \textbf{Low-temperature values undermine the capability of the model to generate plausible code}. A reasonably high level of creativity was demonstrated to be necessary to successfully generate plausible code for some methods. Although, as reported in other studies~\cite{arora2024optimizing,liu2024your,ouyang2023empirical}, low-temperature values guarantee consistency in the responses, they also decrease the spectrum of methods that can be addressed with ChatGPT. This provides new evidence challenging the belief that low-temperature values should be preferred over high values.

\item \textbf{Higher temperature values can generate a higher range of plausible method implementations, with $1.2$ representing a good trade-off between creativity and controllability}. In our experiments, a degree of creativity was necessary to address some methods, that cannot be addressed with lower creativity values. Using the highest creativity levels (i.e., temperature equal to $2$) is not strictly necessary. In fact, a temperature equal to $1.2$ has been sufficient to address the available methods, with a value equal to $2.0$ introducing a small decrease in the percentage of plausible methods generated.

\item \textbf{Small top-p values produce a higher number of plausible methods than higher values}. Our experiments show how small values of top-p are beneficial to the generation of plausible code, with top-p equal to $0.0$ obtaining results that are significantly better than all other configurations. Practitioners should thus consider not using the default values when using ChatGPT for generating code.

\item \textbf{The configuration of the parameters has a major impact on the correctness of the results}. Our study shows how the choice of the parameters may have a relevant impact on the results. In fact, the range of plausible method implementations ranged from $36.4\%$ (for temperature equal to $0.0$ and top-p equal to $0.95$) to $49.3\%$ (for temperature equal to $1.2$ and top-p equal to $0.0$). It is thus important that these parameters are carefully controlled and reported.

\item \textbf{Setting temperature to $0.0$ to diminish non-determinism must be avoided in code generation tasks}. Although setting the temperature to $0.0$ is sometimes used to increase the degree of determinism of ChatGPT, and avoid submitting the same prompts multiple times, this practice should be avoided. First, even with a temperature equal to $0.0$ non-determinism is not fully prevented, and multiple repetitions (e.g., 2 or 3) are needed to properly assess the capability of the model. Moreover, higher temperature values achieved better performance in method generation tasks, and should thus be preferred to using temperature equal to $0.0$.

\item \textbf{Method length and complexity have little impact on the probability of generating plausible implementations}. In the range of the configurations extensively studied (i.e., methods with complexity below 10 and length below 80 locs), the capability of generating plausible configurations has been affected only to a small extent by length and complexity. Developers should thus not have an expectation about the correctness of the code dependent on the length of the code to be generated or its complexity. Moreover, low top-p values did not show any significant dependence on complexity and negligible dependency on length, reinforcing the importance of not using the default values for top-p in code generation tasks. 

\item \textbf{A possibly effective recommendation for the method generation task is using temperature equal to $1.2$, top-p equal to $0.0$, and repeating the requests 5 times}. In this study, we investigated a range of configurations that can be used when querying ChatGPT. Although there are multiple reasonable choices that can produce good results, there is one configuration that performed better than others. Indeed, we cannot consider it a global optimum. It is however a good configuration to start from when employing ChatGPT to generate method implementations: temperature equals $1.2$, top-p equals 0, and completing at least $5$ repetitions.

\end{itemize}

\section{Related Work} \label{sec:related}

This work is mainly related to work on code generation with LLMs and studies about the influence of the parameters that influence LLMs on the correctness of the generated code.

\textbf{Code generation tasks with LLMs}. Code generation tasks assisted by LLMs have been studied according to multiple perspectives~\cite{Liguo:LLM4CG:arXiv:2024,jiang2024surveylargelanguagemodels}. For instance, Corso et al.~\cite{Corso:EmpiricalAssessment:ICPC:2024} compared the effectiveness of multiple AI assistants in method generation tasks, reporting Copilot and ChatGPT as the most effective tools. Yet, only up to one-third of the generated methods were correct.

Research has also examined how the content and structure of prompts affect interactions with AI assistants. For instance, Fagadau et al. investigated the impact of multiple prompts on the generated code~\cite{Fagadau:PromptInfluence:ICPC:2024}.

In addition to studies considering the effectiveness of AI tools in the context of technical tasks, several studies collected data about how developers perceive the utility of these tools. The study by Liang et al.~\cite{liang2024large} about the usability of AI assistants reveals that these tools are already extensively used by practitioners to develop their code, yet there are several areas for improvement.
Similarly, the study by Pinto et al.~\cite{10.1145/3644815.3644949} reports positive feedback by the developers who use these tools. 

LLMs have been reported to be particularly useful to deal with repetitive tasks~\cite{abs-2406-07765}, as well as to ease project onboarding \cite {10662989}, revealing specific use cases where their utility could be maximized. 

Indeed, traces of developers using LLMs, ChatGPT in particular, are extensively available on public repositories. For instance, Tufano et al.~\cite{10.1145/3643991.3644918} analyzed GitHub commits, pull requests, and issues revealing how GitHub developers are using ChatGPT for a number of tasks, including feature implementation and enhancement, documentation, and testing. Broadly speaking, LLMs are already helpful in addressing a number of software engineering tasks, not only code generation~\cite{FanGHLSYZ23}.

This study complements this body of knowledge revealing insights on how to configure and use LLMs, and ChatGPT in particular. In fact, choosing appropriate values for the parameters that control LLMs, and deciding the number of requests to collect for the same prompt, are design decisions that can significantly impact the effectiveness of the interaction with an assistant.

\textbf{Studies about the influence of parameters}. Our study is not the only one considering the impact of the main parameters that affect the output of LLMs in code generation tasks. 

The study by Arora et al.~\cite{arora2024optimizing} investigates the role of temperature and top-p when GPT-3.5-turbo is prompted to generate method implementations, using a dataset of 13 methods. With our study, we extend this preliminary body of evidence by considering GPT-4o, a large-scale dataset with 548 methods, and studying not only the impact of parameters but also the influence of repetitions.  

Liu et al.~\cite{liu2024your} and Ouyang et al.~\cite{ouyang2023empirical} have also studied the effects of these parameters. Their studies show, consistent with our study, that controlling temperature and top-p matter. However, they recommend limiting creativity and randomness by using low values of both temperature and top-p. This is in contrast with our findings. Our analysis of ChatGPT's behavior with repeated submissions reveals that creativity facilitates the generation of challenging implementations. If it is true that low-temperature values may produce more consistent results, this may also limit the range of implementations that can be obtained. In fact, we ultimately recommend using five repetitions with temperature values greater or equal to $1.2$, rather than using low-temperature values that may limit the effectiveness of the ChatGPT.

\section{Conclusion} \label{sec:conclusions}
Software development is increasingly conceived as a collaboration activity between developers and AIs. Indeed, IDEs already implement features to enable interactive development, with AI suggesting implementations that are reused by developers.

Although multiple studies show this interaction can be successful, there is still limited understanding of how the models must be configured and used in the context of code generation tasks. This study addresses this gap, systematically investigating the impact of several key parameters, including the repeated submission of a prompt to accommodate for the non-deterministic nature of the models.

Our study reveals several key findings about the usage of ChatGPT. In particular, we discovered how creativity, although up to a limited extent, is useful to increase the range of methods whose code can be generated correctly. A major role is played by parameter top-p, which is commonly underrated, and instead has a major impact on the correctness of the results, with lower values producing better results. Finally, prompts should be submitted multiple times, with $5$ repetitions combined with a temperature of $1.2$ resulting in an effective configuration in our experiments.  

Future work concerns two main research directions. One is about replicating this experiment with other AI assistants, to validate our findings in multiple contexts. The second research direction concerns finding strategies to deal with the need to submit the same prompt multiple times to obtain a useful result, and thus developing approaches able to select or merge multiple responses automatically. 

\section*{Acknowledgment}
This work was partially supported by the MUR under the grant “Dipartimenti di Eccellenza 2023-2027" of the Department of Informatics, Systems and Communication of the University of Milano-Bicocca, Italy; and by the Engineered MachinE Learning-intensive IoT systems (EMELIOT) national research project, which has been funded by the MUR under the PRIN 2020 program (Contract 2020W3A5FY).

\bibliographystyle{IEEEtran}
  \bibliography{main}

\vspace{12pt}

\end{document}